\documentclass[aps,prl,twocolumn,floatfix]{revtex4}
\usepackage{graphicx}
\begin{document}

\title{Inhomogeneity--induced enhancement of the pairing interaction in cuprates}
\author {Maciej M. Ma\'ska}
\author {\.Zaneta \'Sled\'z}
\author {Katarzyna Czajka}
\author{Marcin Mierzejewski}
\affiliation{Institute of Physics, University of Silesia,
 40-007 Katowice, Poland}
\date{\today}

\begin{abstract}
Scanning tunneling spectroscopy  has recently discovered a positive correlation 
between the magnitude of the superconducting gap and positions of dopant oxygen atoms in Bi-based
cuprates. We propose a microscopic mechanism that could be responsible for this  effect. 
In particular, we demonstrate that the dopant--induced spatial variation
of the atomic levels
always enhances the superexchange interaction.  
\end{abstract}


\maketitle

Scanning tunneling microscopy (STM) and scanning tunneling spectroscopy (STS) 
have recently confirmed that nanoscale electronic inhomogeneity is 
an inherent feature of many groups of high--temperature superconductors (HTSC).
Being capable of direct probing the local density of states 
(LDOS), these methods revealed strong spatial modulation of the energy gap 
in Bi--based compounds \cite{Pan}. This modulation occurs on a very short length 
scale, of 
the order of the coherence length. It has been shown that 
the system consists of regions of relatively small gap ($\Delta \approx 25\div 35 
{\rm meV}$) with high and sharp coherence peaks and 
regions of larger gap ($\Delta \approx 50\div 75 {\rm meV}$) with small and 
broad peaks \cite{Lang}. 
Since its discovery, the inhomogeneity was commonly attributed
to a disorder introduced by poorly screened electrostatic potential of 
the out--of--plane oxygen dopant atoms. A lot of theoretical works exploited
this idea, predicting modification of the energy gap in the vicinity of
a dopant atom \cite{wang}.
Very recently, STM experiments have shown a strong
correlation between position of the dopant atoms and all manifestations
of the nanoscale electronic disorder \cite{McElroy,mashima}. Thus, these 
experiments proved the impurities to be the source of the inhomogeneity. 
On the other hand, they revealed a very important feature: there is a {\em positive} 
correlation between the position of a dopant atom and the magnitude of a gap \cite{McElroy}.
The sign of this correlation function contradicts the previous theoretical 
predictions, based on a direct reduction of the gap by the dopant's 
electrostatic potential \cite{wang}. Therefore, a new mechanism that is capable of
gap enhancing close to  impurities, is needed to explain the correlations.
It was shown by Nunner {\em et al.} \cite{nunner}, that an assumption 
of an enhancement of the pairing potential by the presence of a nearby dopant atom
%
leads to the correct correlation between the height of the coherence peaks and the 
magnitude of the gap as well as between the position of the dopant 
atom and the magnitude of the gap.  The spectral properties could also 
be explained in a different way, assuming that the observed peaks at the edges
of the gap arise from resonant bound states rather than they are the coherence peaks 
\cite{kapitulnik}.

From the theoretical point of view,  various parameters of the microscopic models of HTSC
can be modified by the presence
of the dopant atoms. 
 These atoms are charged impurities and since cuprates
are close to the insulating state, the number of carriers is too low to effectively 
screen their electrostatic potential. As a result, atomic levels in the CuO$_2$ plane 
are shifted in the vicinity of the dopants. 
The presence of the dopants also induces a local distortion of the lattice \cite{nun1}. 
This, in turn, may modify the hopping matrix elements as well as the electron--phonon coupling.
It has recently been shown that inhomogeneity itself can increase the superconducting 
transition temperature \cite{kiv1,kiv2}.


Here, we show that in strongly correlated systems, the position--dependent shift of the atomic 
levels  alone is sufficient to enhance the pairing 
interaction, thereby leading to the correct sign of the dopant--gap correlation 
function. 
We assume that HTSC can be described by the $t$--$J$ model, 
with the exchange 
interaction as the main pairing mechanism \cite{anderson}. The physical picture behind 
our analysis 
is based on the same arguments, which allow one to derive the $t$--$J$ model from 
the Hubbard one \cite{spalek}. Since strong Coulomb repulsion excludes double 
occupancy, eigenstates of the half--filled Hubbard model are built out of localized
electron states, with one electron per lattice site. A virtual hopping of an 
electron to a neighboring site (occupied by an electron with the opposite spin) 
gives rise to an effective spin exchange interaction with potential $J=4 t^2/U$, 
where $t$ is the hopping integral and $U$ is the Coulomb potential. Obviously, 
local modification of the hopping matrix elements
leads either to an enhancement \cite{zhu} or to a reduction of the exchange interaction.
Recent density functional investigations indicate that the dopant oxygen atoms are 
responsible for displacements of atoms in the CuO$_2$ plane  \cite{nun1}. 
 Since these displacements are perpendicular to the plane,
such a distortion increases the interatomic distances. Therefore, one might expect
a reduction of the hopping integrals and a simultaneous decrease of $J$.    
However, the effective exchange interaction should depend also on
the atomic levels at sites, which are involved in the virtual hopping. 
It comes from the fact, that the energy of the virtual doubly 
occupied state depends on these levels.
The basic aim of the present manuscript is to show that
the diagonal disorder always leads to
an enhancement of the antiferromagnetic exchange interaction.
As a result the superconducting gap increases in the vicinity of the dopant
atoms.

We start with the two--dimensional one band Hubbard model on a square lattice
\begin{eqnarray}
H&=&-t \sum_{\langle i,j \rangle \sigma} 
\left( a^{\dagger}_{i \sigma}a_{j \sigma} + {\rm h.c.} \right)
 + \sum_{i \sigma} (V_i-\mu)  a^{\dagger}_{i \sigma}a_{i \sigma} \nonumber \\
&&+\;U \sum_{i} a^{\dagger}_{i \uparrow}a_{i \uparrow}
a^{\dagger}_{i \downarrow}a_{i \downarrow},
\end{eqnarray}
where $a^{\dagger}_{i \sigma}$ creates an electron with spin  $\sigma$ at site $i$,
$\mu$  is the chemical potential and $V_i$ is the atomic level 
at site $i$. The latter quantity accounts  for the electrostatic potential of 
the dopant atoms. 

In the following we argue that the enhancement of the pairing 
interaction is driven by a random distribution of $V_i$. However,
there is a severe drawback of such assumption: strong enough dopant's potential
should significantly affects the electron concentration in its neighborhood, 
resulting in a highly inhomogeneous charge distribution. On the other hand, the
STS experiments show that the root mean square of the electron density in 
these systems are below 10\% \cite{McElroy}. One can expect however, that strong Coulomb 
repulsion along with vicinity to half--filling, meaningfully reduce the 
dopant--induced inhomogeneity. Namely, the decrease of the local density of 
electrons close to the dopant atoms must result in its increase away from them. 
Such an increase is energetically highly unfavorable. In particular, at 
half--filling and for a strong Coulomb repulsion impurities cannot produce 
any charge inhomogeneities at all.  In order to verify the
effectiveness of this mechanism for a physically relevant doping, 
we have performed exact diagonalization study on a finite--size cluster.  
We have considered two impurities located $z=1.5a$ above the cluster with
$a$ being the lattice constant.  
It is the distance between the dopant oxygen atom and the CuO$_2$ plane
in Bi--based HTSC. These impurities act as a source of a screened electrostatic
potential, that shifts the atomic levels
$V_i=V_0 \sum_m \exp(-R_{mi}/\lambda)/ \tilde{R}_{mi}$, where the summation
is carried out over all dopant atoms.
Here, $R_{mi}$ is a distance between site $i$ and site $m$ (above which the dopant atom 
is located) and $\tilde{R}_{mi} =\sqrt{R_{mi}^2+z^2}$.
Such a form of $V_i$ accounts for the fact that the potential 
is screened only by electrons moving in the CuO$_2$ plane.
However, our final conclusions are independent of the form of $V_i$. 


\begin{figure}[h]
\includegraphics[width=8cm]{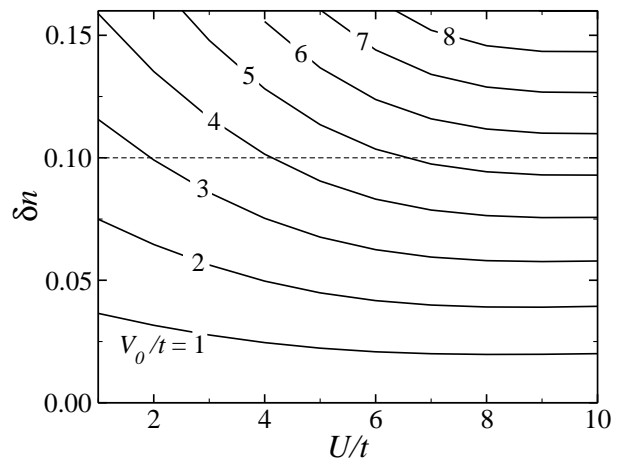}
\caption{Root mean square of the local electron density as a function of $U$ for various values of $V_0$,
as explicitly indicated in the figure. 
Results have been obtained for 10 electrons on a 12-site cluster 
with periodic boundary conditions. $\lambda$ equals two lattice constants.}
\label{fig1}
\end{figure}

Fig. \ref{fig1} presents root mean square of the local electron concentration
$\delta n$ as a function of the Coulomb interaction $U$ for various values of $V_0$. 
One can note a significant reduction of the charge inhomogeneity in 
a large--$U$ regime in spite of the strong variation of the electrostatic potential 
$V_i$. 
In the uncorrelated case, inhomogeneity of the charge distribution depends on the
ratio $V_0/t$. Contrary to this, in the strongly correlated nearly half--filled system
inhomogeneity is determined predominantly by $V_0/U$. Therefore,  small charge density
variation observed in STS experiments does not require weak electrostatic potential of the dopant atoms.

We now turn to the main problem, i.e., the enhancement of the pairing interaction 
by the dopant atoms.
For this sake, we make use of the Hubbard operators 
$X^{\alpha, \beta}_{i}=|\alpha\rangle_i \langle \beta |_i$
($\alpha,\beta=0,\uparrow,\downarrow,\uparrow\downarrow$)  and carry out 
the canonical transformation 
$H \rightarrow \exp(-S) H \exp(S)$. The generator $S=-S^{\dagger}$  
is chosen in such a way that the transformation eliminates the transfer of electrons
between lower and upper Hubbard subbands. Straightforward calculations lead to the 
generator of the form
\begin{equation}
S=\sum_{\langle i,j \rangle,\sigma}
 s(\sigma) t \left[ 
\frac{X^{\uparrow\downarrow,\bar{\sigma}}_i X^{0,\sigma}_j}{U+\left(V_i-V_j \right)}
-
\frac{X^{\sigma,0}_i X^{\bar{\sigma},\uparrow\downarrow}_j}{U-\left( V_i-V_j \right)}
\right],
\end{equation}
where $s(\uparrow)=1$ and $s(\downarrow)=-1$. 

Next, one can project out states with doubly occupied sites and obtain the effective
Hamiltonian acting in the space spanned by states $|0 \rangle_i$, 
$|\uparrow \rangle_i$ and  $|\downarrow \rangle_i$. This is a generalized $t$--$J$ 
Hamiltonian with a site dependent spin exchange interaction:
\begin{eqnarray}
H_{t-J}&=&-t \sum_{\langle i,j \rangle \sigma} 
 \left( \tilde{a}^{\dagger}_{i \sigma}\tilde{a}_{j \sigma} + {\rm h.c.} \right) 
 + \sum_{i \sigma} (V_i-\mu)  \tilde{a}^{\dagger}_{i \sigma} \tilde{a}_{i \sigma} \nonumber \\
&&+ \sum_{\langle i,j \rangle} J_{ij} \left(
\vec{S}_i \cdot \vec{S}_j-\frac{1}{4} n_i n_j \right).
\label{tj}
\end{eqnarray}
Here, $\tilde{a}^\dagger_{i \sigma}$  creates an electron at site $i$ 
if this site previously had no electron. The interaction $J_{ij}$
 is given by


\begin{equation}
J_{ij}=\frac{4 t^2}{U}\:\left(1+\eta_{ij}\right),
\label{J}
\end{equation}
where
\begin{equation}
\eta_{ij}= \frac{(V_i-V_j)^2}{U^2-(V_i-V_j)^2} \ge 0.
\label{JJ}
\end{equation}

\begin{figure}[h]
\includegraphics[width=3.7cm]{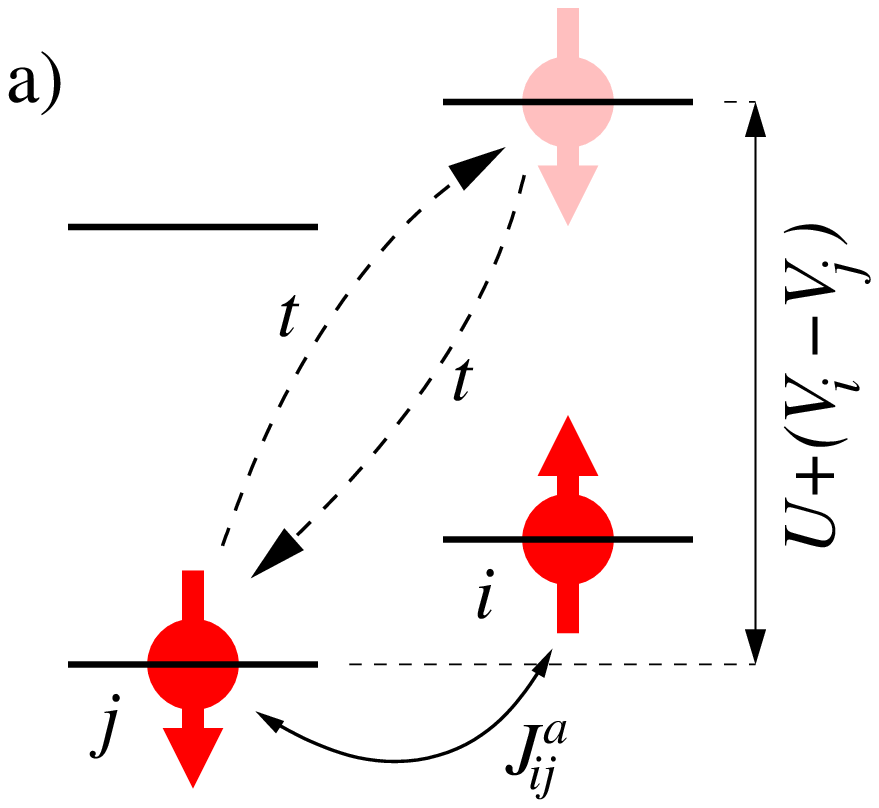}\hspace*{3mm} \includegraphics[width=3.7cm]{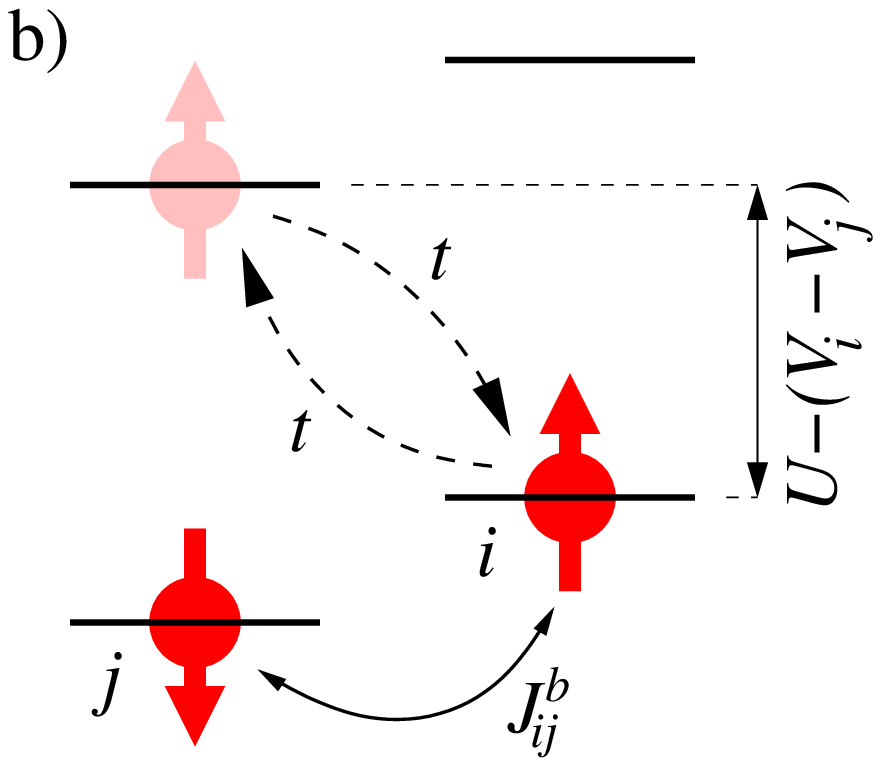}
\caption{(Color online) Second order two--site processes contributing the effective coupling
between spins at sites $i$ and $j$.}
\label{fig2}
\end{figure}

In the homogeneous case ($V_i=$ const), $J_{ij}$ reduces itself
to the standard form $J_0=4t^2/U$. Otherwise, it
is always {\em larger} than $J_0$, independently of the distribution of the atomic levels and
any particular form of $V_i$.
Moreover, the strength of the coupling increases with the difference between
the atomic levels at neighboring sites. Assuming $J_{ij}$ to be the effective
pairing interaction one comes to the conclusion that superconductivity can be
enhanced by inhomogeneous distribution of the atomic levels. Since
these levels directly affect the pairing interaction, spatial
variation of the superconducting gap does not
require  strongly  inhomogeneous charge distribution.
The surprising enhancement
can be easily understood from the analysis of virtual processes presented in Fig.
\ref{fig2}.
The resulting exchange interaction is proportional to the squared hopping matrix elements ($t^2$)
and inversely proportional to the energy of the intermediate state with doubly
occupied site.
There are two second order processes that
lead to the effective antiferromagnetic coupling of spins at sites $i$ and $j$. 
The    
first of them (Fig. \ref{fig2}a) gives $J^a_{ij}=2t^2/[U+(V_i-V_j)]$, whereas the 
second one (Fig. \ref{fig2}b) gives $J^b_{ij}=2t^2/[U-(V_i-V_j)]$. Taking into account 
both these processes one obtains $J_{ij}= J^a_{ij}+J^b_{ij}$, what is the value 
given by Eq. \ref{J}.

Since the Coulomb repulsion is the dominating energy scale in cuprates, one can assume that  $|V_i-V_j| \ll U $.
Then, an approximate formula $\eta_{ij} \simeq (V_i-V_j)^2/U^2$ holds and
one can see that  $\eta_{ij}$ decreases much faster than $V_i$ when the distance from
the dopant atom increases.  
As a result, a significant enhancement
of the exchange interaction is expected only in the closest vicinity of the dopant. 

It is known that an enhancement of the pairing potential close to impurities
leads to the correct sign of the impurity--gap correlation function as it has been shown
in Refs. \cite{nunner,corfun}. 
However, in these papers the spatial variation of the pairing potential is of phenomenological
origin. Contrary to this, we have proposed a microscopic mechanism that gives an explicit form of
this dependence.   
In the following, we investigate whether the proposed scenario
quantitatively reproduces the STS data. In particular, it is necessary to check 
whether a significant enhancement of the superconducting gap can be obtained for the model parameters, which
do not lead to strong fluctuations of the electron concentration.


In order to estimate the spatial variation of the superconducting order parameter,
we have assumed the  resonating valence bond (RVB) scenario \cite{anderson} and
investigated the Hamiltonian (\ref{tj}) in the mean--field 
approximation with renormalized hopping integral $t \rightarrow \tilde{t}= (1-n) t $ 
\cite{anderson,kotliar}. Within such an approach the constraint of no double occupancy 
is not fulfilled exactly. On the other hand, it is a method that allows one to investigate
superconductivity in inhomogeneous systems by means of the Bogoliubov--de Gennes (BdG)
equations. Following the experimental results we restrict further study  to such values
of the model parameters $  U$ and $ V_0$, which give small electron density fluctuations $\delta n < 0.1 $ 
(see Fig.~\ref{fig1}).  
Therefore, we neglect the term containing $V_i$ in the Hamiltonian (\ref{tj}), that
is responsible for the charge inhomogeneity. Since the mean--field RVB approach leads
to  an incorrect doping dependence of the critical temperature, this term could be responsible
for an unphysical contribution to the spatial dependence of the order parameter.  

The mean--field  Hamiltonian takes on the form:
\begin{eqnarray}
H_{t-J}&=&-\tilde{t} \sum_{\langle i,j \rangle \sigma} 
 \left( a^{\dagger}_{i \sigma} a_{j \sigma} + {\rm h.c.} \right) 
 -\mu \sum_{i \sigma} a^{\dagger}_{i \sigma} a_{i \sigma} \nonumber \\
&&+ \sum_{\langle i,j \rangle} \left[
\Delta_{ij} \left(
a^{\dagger}_{i \uparrow}a^{\dagger}_{j \downarrow}+
a^{\dagger}_{j \uparrow}a^{\dagger}_{i \downarrow}
\right)+ {\rm h.c.} \right], 
\label{tjmf}
\end{eqnarray}
where $\Delta_{ij}=-J_{ij}\langle a_{i \downarrow} a_{j \uparrow}+
 a_{j \downarrow} a_{i \uparrow} \rangle/2 $.
It can be diagonalized with the help of
the transformation:
\begin{equation}
 c_{i \sigma}= \sum_{n} ( u_{in} \gamma_{n \sigma}
-s(\sigma) v^*_{in}\gamma^{\dagger}_{n \bar{\sigma}}),
\end{equation}
where $u_{in}$ and $v_{in}$ fulfill the BdG equations:
\begin{equation}
\sum_j
\left(
\begin{array}{cc}
{\cal H}_{ij} & \Delta_{ij} \delta_{\langle ij \rangle} \\
\Delta^*_{ij} \delta_{\langle ij \rangle}  & - {\cal H}_{ij}
\end{array}
\right)
\left(
\begin{array}{c}
u_{jn} \\
v_{jn}
\end{array}
\right)
= \varepsilon_n
\left(
\begin{array}{c}
u_{in} \\
v_{in}
\end{array}
\right),
\end{equation}
with ${\cal H}_{ij}= -\tilde{t}  \delta_{\langle ij \rangle} - \mu  \delta_{ij} $.
Here, $ \delta_{\langle ij \rangle}=1$ for the neighboring sites $i,j$ and $0$ otherwise.
The superconducting order parameter is determined self-consistently by:
\begin{equation}
\Delta_{ij}=\frac{J_{ij}}{2} \sum_{n} 
\left( u_{in} v^*_{jn}+u_{jn}v^*_{in} \right) \tanh \frac{\varepsilon_n}{2 k_{\rm B}T}. 
\end{equation}
              
\begin{figure}[h]
\includegraphics[width=7cm]{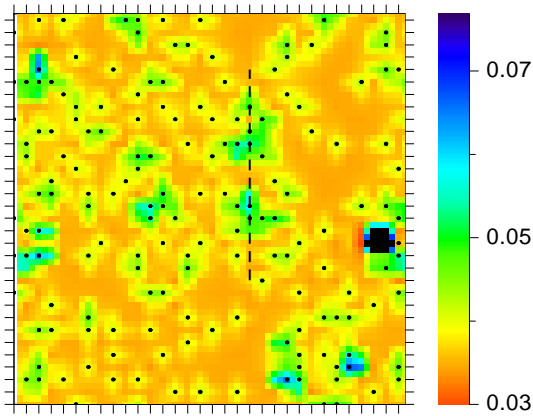}
\caption{(Color online) $\Delta_i/t$ calculated for $U=8t$, $V_0=5t$, $\lambda=2a$ and $\bar{n} \simeq 0.83$
The positions of impurities are marked by dots. The dashed line shows the cross--section, along which
the LDOS presented in Fig. \ref{fig4} was calculated.
}
\label{fig3}
\end{figure} 

\begin{figure}[h]
\includegraphics[width=5cm]{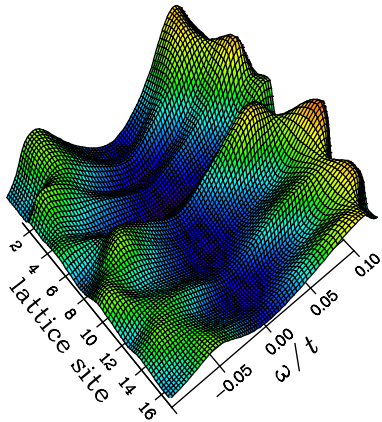}
\caption{(Color online) LDOS calculated along the line marked in Fig. \ref{fig3} for the same model parameters.}
\label{fig4}
\end{figure}

We have solved the BdG equations for the $d$--wave symmetry. The calculations have been carried 
out on a
$32\times 32$ cluster with 170 randomly distributed impurities, what gives concentration $16\%$. Assuming that 
each dopant introduces one hole into the CuO$_2$ plane, we have performed calculations for the same hole 
concentration. 
Fig. \ref{fig3} shows the spatial variation of the superconducting order parameter 
obtained for $U=8t$, $V_0=5t$ and $\lambda=2a$.
For such model parameters exact diagonalization 
study gives the root mean square of the electron density below 0.1. 
(see Fig. \ref{fig1}), what is comparable to the experimentally observed fluctuations 
\cite{McElroy}.
One can note a significant spatial variation of the superconducting gap.
We have found that the root mean square of the order parameter
equals  25\% of its average value.  
As expected $\Delta_i$ increases 
in the vicinity of each impurity. However, the degree of this enhancement 
strongly depends
on a configuration of impurities. It is significant in the regions, where 
several impurities are located close to each other. Otherwise it is much weaker.
This result follows from the specific form of exchange interaction as given by Eq. (\ref{J}).
Due to the screening of the electrostatic potential, $\eta_{ij}$ gives nonnegligible contribution
to $J_{ij}$ only in the closest neighborhood of an impurity. 
Consequently, the pairing interaction is enhanced in a region larger than the coherence length
only provided this region contains several impurities.  Figs. \ref{fig3} and \ref{fig4} show that the 
mean--field approach correctly reproduces
also the main qualitative features of the local density of states (LDOS).     
In particular, increase of the superconducting gap in the vicinity of impurities
is accompanied by a reduction  of the height of the coherence peaks. The reduction is however
much weaker than observed experimentally. This discrepancy can be attributed either
to the limitations of the mean--field analysis or/and to fact, that we have considered
only the increase of the pairing interaction and neglected other possible
dopant--induced effects, e.g., modulation of the hopping integral.

To conclude, we have derived a purely microscopic mechanism that can be responsible for 
the observed enhancement of the superconducting gap in the vicinity of impurities. 
Our approach follows
the generally accepted view that high temperature superconductivity originates from
the exchange interaction between electrons in the lower Hubbard subband. 
We have considered a simple scenario of the influence of 
dopant atoms on electrons moving in the CuO$_2$ planes. 
It is based on the shift of the atomic levels by the dopant's electrostatic potential.
One of the most attractive features of this approach concerns
its simplicity. On the other, hand such a simple approach neglects other possible effects 
originating from the presence of the dopant oxygen atoms. In particular, we expect that 
vibrations of these atoms can further reduce the height of the coherence peaks in the large gap
areas, as observed experimentally.

\acknowledgements
This work has been supported by the Polish Ministry of Education and Science
under Grants No. 1~P03B~071~30.


\begin{thebibliography}{99}
\bibitem{Pan} S.H. Pan {\em et al.}, Nature {\bf 413}, 282 (2001).
\bibitem{Lang} K.M. Lang {\em et al.}, Nature {\bf 415}, 412 (2001).
\bibitem{wang} Z. Wang {\em et al.}, Phys. Rev. B {\bf 65}, 064509 (2002);
Q.--H. Wang {\em et al.}, Phys. Rev. B {\bf 65}, 054501 (2001);
I. Martin and A.V. Balatski, Physica C {\bf 357--360}, 46 (2001);
A.V. Balatsky {\em et al.}, Rev. Mod. Phys. {\bf 78}, 373 (2006).
\bibitem{McElroy} K. McElroy {\em et al.}, Science {\bf 309}, 1048 (2005).
\bibitem{mashima} H. Mashima {\em et al.}, Phys. Rev. B {\bf 73}, 060502 (2006).
\bibitem{nunner} T.S. Nunner {\em et al.}, Phys. Rev. Lett. {\bf 95}, 177003 (2005). 
\bibitem{kapitulnik} A.C. Feng {\em et al.}, Phys. Rev. Lett. {\bf 96}, 017007 (2006).
\bibitem{nun1} Y. He {\em et al.}, Phys. Rev. Lett. {\bf 96}, 197002 (2006). 
\bibitem{kiv1} I. Martin {\em et al.}, Phys. Rev. B {\bf 72}, 060502 (2005);
K.~Aryanpour {\em et al.}, Phys. Rev. B {\bf 73}, 104518 (2006).
\bibitem{kiv2} W.--F. Tsai and S. Kivelson, Phys. Rev. B {\bf 73}, 214510 (2006).
\bibitem{anderson} G. Baskaran {\em et al.}, Solid State Commun. {\bf 63}, 973 (1987).
\bibitem{spalek} A. Chao {\em et al.}, J. Phys. C {\bf 10}, L271 (1977);Phys. Rev. B {\bf 18}, 3453 (1978).
\bibitem{zhu} J.--X. Zhu, {\tt cond-mat/0508646}.
\bibitem{corfun} T.S. Nunner {\em et al.}, {\tt cond-mat/0606685}.
\bibitem{kotliar} G. Kotliar, Phys. Rev. B {\bf 37} 3664 (1988).

\end{thebibliography}
\end{document}